\documentclass[10pt,journal,compsoc]{IEEEtran}

\newif\ifpdf
\ifx\pdfoutput\undefined
   \pdffalse
\else
   \pdfoutput=1
   \pdftrue
\fi
\ifpdf
   \usepackage{graphicx}
   \usepackage{epstopdf}
\else
   \usepackage{graphicx}
\fi

\usepackage[T1]{fontenc}
\usepackage[utf8]{inputenc}

\usepackage{combelow}
\usepackage{newunicodechar}
\usepackage[noend]{algpseudocode}
\usepackage{cite}
\usepackage{balance}
\usepackage{siunitx}
\usepackage{upgreek}
\usepackage{graphicx}
\usepackage{commath}
\usepackage{color}
\usepackage{array}
\usepackage{float}
\usepackage{longtable}
\usepackage{algorithmicx}
\usepackage{algpseudocode}
\usepackage{algorithm}
\usepackage{amsmath}
\usepackage{amssymb}
\usepackage{multicol}

\begin{document}
\title{Comparison of wide-band vibrotactile \\ and friction modulation surface gratings}

\author{Roman~V.~Grigorii,~\IEEEmembership{Member,~IEEE,}
        Yifei Li,~\IEEEmembership{Non-member,~IEEE,}
        Michael A. Peshkin,~\IEEEmembership{Senior Member,~IEEE,}
        J.~Edward~Colgate,~\IEEEmembership{Fellow,~IEEE,}
        
\IEEEcompsocitemizethanks{\IEEEcompsocthanksitem The authors are with the Department of Mechanical Engineering in Northwestern University, Evanston, IL 60208. Correspondence e-mail with Roman Grigorii: romangrigorii2015@u.northwestern.edu}
\thanks{Manuscript received x; revised x}}

\markboth{Journal of \LaTeX\ Class Files,~Vol.~x, No.~x, January~2021}
{Shell \MakeLowercase{\textit{et al.}}: Bare Demo of IEEEtran.cls for Computer Society Journals}

\IEEEtitleabstractindextext{

\begin{abstract}

This study seeks to understand conditions under which virtual gratings produced via vibrotaction and friction modulation are perceived as similar and to find physical origins in the results. To accomplish this, we developed two single-axis devices, one based on electroadhesion and one based on out-of-plane vibration. The two devices had identical touch surfaces, and the vibrotactile device used a novel closed-loop controller to achieve precise control of out-of-plane plate displacement under varying load conditions across a wide range of frequencies. A first study measured the perceptual intensity equivalence curve of gratings generated under electroadhesion and vibrotaction across the 20-400Hz frequency range. A second study assessed the perceptual similarity between two forms of skin excitation given the same driving frequency and same perceived intensity. Our results indicate that it is largely the out-of-plane velocity that predicts vibrotactile intensity relative to shear forces generated by friction modulation. A high degree of perceptual similarity between gratings generated through friction modulation and through vibrotaction is apparent and tends to scale with actuation frequency suggesting perceptual indifference to the manner of fingerpad actuation in the upper frequency range. 

\end{abstract}

}

\maketitle
\IEEEdisplaynontitleabstractindextext
\IEEEpeerreviewmaketitle
\IEEEraisesectionheading{\section{Introduction}\label{sec:introduction}}

For decades, vibrotaction (VT) has served as a cost-effective and robust tool for conveying meaningful tactile information in electronic devices. Due to the remarkably low perceptual threshold to vibration among humans ($\approx$ 0.1$\mu$m at 250 Hz \cite{verrillo1992vibration}), this form of stimulation has proved to be highly versatile in generating feedback and enriching tactile user experience in a number of haptic applications \cite{strohmeier2017generating, lieberman2007tikl,zheng2010vibrotactile}. While VT technology is ubiquitous today, the breadth of its rendering potential is limited due to issues associated with vibrating a mass. Among these technological limitations are: narrow actuation bandwidth, inability to produce sustained forces at the periphery, and difficulty spatially isolating the stimulus in multi-touch user interaction without the need for complex workarounds \cite{hudin2013}. As a consequence of these factors, the rendering of wide-bandwidth surface textures, virtual switches, detents, and shapes that are spatially distributed across the haptic surface, are fundamentally difficult to achieve through VT alone \cite{allerkamp2007vibrotactile}. 

Tactile feedback provided by varying friction under the fingertip, termed friction modulation (FM), has gained increasing interest within the haptics community in the past decade (although the underlying technologies date to early-mid 20$^{\text{th}}$ century \cite{johnsen1923physical,salbu1964compressible}). One is termed ultrasonic vibration and functions by levitating the skin of the fingertip by bouncing it on a high-pressure cushion of air that is trapped between the skin and an ultrasonically vibrating surface \cite{wiertlewski2016partial} \cite{giraud2007}. The second method operates by generating electrostatic charge on an electrode surface, which creates electrostatic forces that pull on the skin of the contacting finger to create broader asperity contacts and, as a consequence, increased friction. When electrostatic forces are applied at ultrasonic frequencies (``electroadhesion'') and then amplitude modulated at a perceptible frequency, the result is  a more stable and wider-bandwidth tactile effect than if the electrostatic charge is applied directly at the perceptible frequencies (``electrovibration'') \cite{Shultz2018}. Amplitude modulation of an ultrasonic carrier, be it vibratory or electrostatic, provides a much wider bandwidth of tactile stimulation than VT, enabling an overall richer rendering gamut.

FM based tactile feedback has been utilized in precise playback of high-bandwidth forces arising from finger interacting with natural textures \cite{grigorii2020closed}; producing active lateral forces that create realistic renderings of button click sensation through active force feedback on a flat surface \cite{heng2018}; modulating a surface's static coefficient of friction and thereby its perceived stickiness \cite{grigorii2019}; and generating switches and shapes \cite{teslatouch2009,kim2013tactile}. Besides offering a wide range of tactile effects, the solid state nature of electroadhesion ensures that it is relatively power efficient and scalable to a range of screen sizes, shapes, and even material types (as long as the material surface consists of an insulated conductor). In addition, by patterning electrodes, it is possible to create tactile surfaces that support multi-touch interaction \cite{ilkhani2018creating}.

Although electroadhesion-based FM is a promising alternative to VT from technological standpoint, it is not yet understood what the perceptual trade-offs are when one method is selected over the other to create effects such as tactile gratings and textures. To date, haptics research has demonstrated that VT can serve as a tool of to convey complex textural qualities \cite{allerkamp2007vibrotactile, strohmeier2017generating, okamura1998vibration} and a similar narrative has developed among researchers utilizing FM technology \cite{meyer2015modeling,saleem2019tactile,bernard2018harmonious,friesen2020building}. There is also some  evidence that the there may be perceptual benefit in combining the two methods \cite{ito2019tactile}; however, no systematic effort has been made to relate the two stimulation types perceptually or physically.

In this work, we begin to make such an effort by developing a surface haptic device capable of producing a controlled VT stimulus across a wide range of frequencies and amplitudes. Our apparatus can display both low-frequency, high-amplitude displacement and high-frequency vibration. Using this device and an electroadhesive FM device having the identical surface shape and texture, we perform two psychophysical studies that perceptually compare VT and FM virtual gratings and furthermore tie these results to known physical and physiological properties of the fingertip. Our experiments involve gratings rendered in time rather than space which offers straightforward dynamic analyses; however, the results should translate to spatial gratings when corrected for finger velocity.

\subsection{Physics and perception of FM and VT stimulation}

The physics underlying the FM and VT approaches are starkly different. In the case of FM, the input is a force \cite{contactmechanicsbopersson,shultz2015surface} and in the case of VT, the input is a displacement \cite{wiertlewskimechanical,hajian1997}. Not only are the causalities opposite, but the directions of stimulation are orthogonal to each other: FM produces lateral forces while VT produces (in our case) normal displacements. How might these physical differences affect perception?  

Dynamic tissue deformation around each mechanoreceptor is what fundamentally predicts resulting mechanoreceptor response, and therefore should be the dominant factor that correlates with perceived stimulus intensity \cite{roudaut2012touch}. It stands to reason that two stimulation methods will be matched in intensity when the overall deformation at the periphery is matched.  However, since the VT and FM methods act in different directions, a perfect perceptual match may never occur. The consequence of differences in stimulus direction on the mechanoreceptor response has been previously addressed \cite{birznieks2001encoding}.  This study and others suggest that low frequency stimuli which may be encoded by SA-I, SA-II, and FA-I \cite{johansson1976skin} cutaneous afferents as well as those effecting kinesthetic sense \cite{klatzky2003touch,collins2005cutaneous} is direction dependent, while high frequency ($>$40Hz) stimuli encoded by the PC afferent may not be \cite{abraira2013sensory}. It is thus anticipated that perceptual differences between VT and FM will emerge at low stimulation frequencies where two fundamentally different and direction-sensitive afferents will be engaged to a variable degree, but that a perceptual match will be feasible at higher frequencies where intensity is encoded by direction insensitive PC afferents. 

With this insight in mind, the goal of this work is to answer two questions:

$\newline$ 1) What physical quantity (e.g., skin displacement, velocity, applied force, etc.) best predicts the relative intensity of FM and VT?

$\newline$ 2) How perceptually similar are the two forms of excitation when their overall intensity has been matched?

\subsection{Precision in tactile stimulation}

One challenge in generating surface haptic feedback is achieving precise control of the stimulus under active touch. Uncertainties, dynamics, and disturbances originate in both the display device and the human, the latter including factors such as normal load, incidence angle, and swipe speed \cite{meyer2014dynamics}. On the device side, there are known issues in controlling friction forces \cite{grigorii2020closed} and vibration amplitude \cite{wiertleski2011}. An important consideration in vibrotaction is resonant frequency, which corresponds to the point of lowest system impedance and thus the point of largest sensitivity to external loads. In addition, the dynamic phase shift at resonance limits direct application of closed loop control. One can widen overall system bandwidth by shifting resonant frequency upward via increased stiffness, but this is done at the cost of range of motion \cite{wiertleski2011}. The mass of moving components can also be reduced, but this approach is typically limited by size and material constraints. 

To overcome these limitations for VT feedback, we developed a novel vibrotactile device and control method. The device consisted of a section of 3M surface capacitive touch screen (identical to that used for FM) mounted on two voice coil actuators providing about 1mm of vertical displacement.  The substrate mass and actuator stiffness yielded a resonant frequency of 190Hz. Despite this relatively low value, the feedback controller described in detail in the next section enabled a precise tactile stimulation profile across the surface for a large range of amplitudes (1$\mu$m - 1mm) and frequencies (20Hz - 400Hz).

\section{Methods}
\subsection{Apparatus}

\begin{figure}[t]
\includegraphics[width=8.5 cm,keepaspectratio]{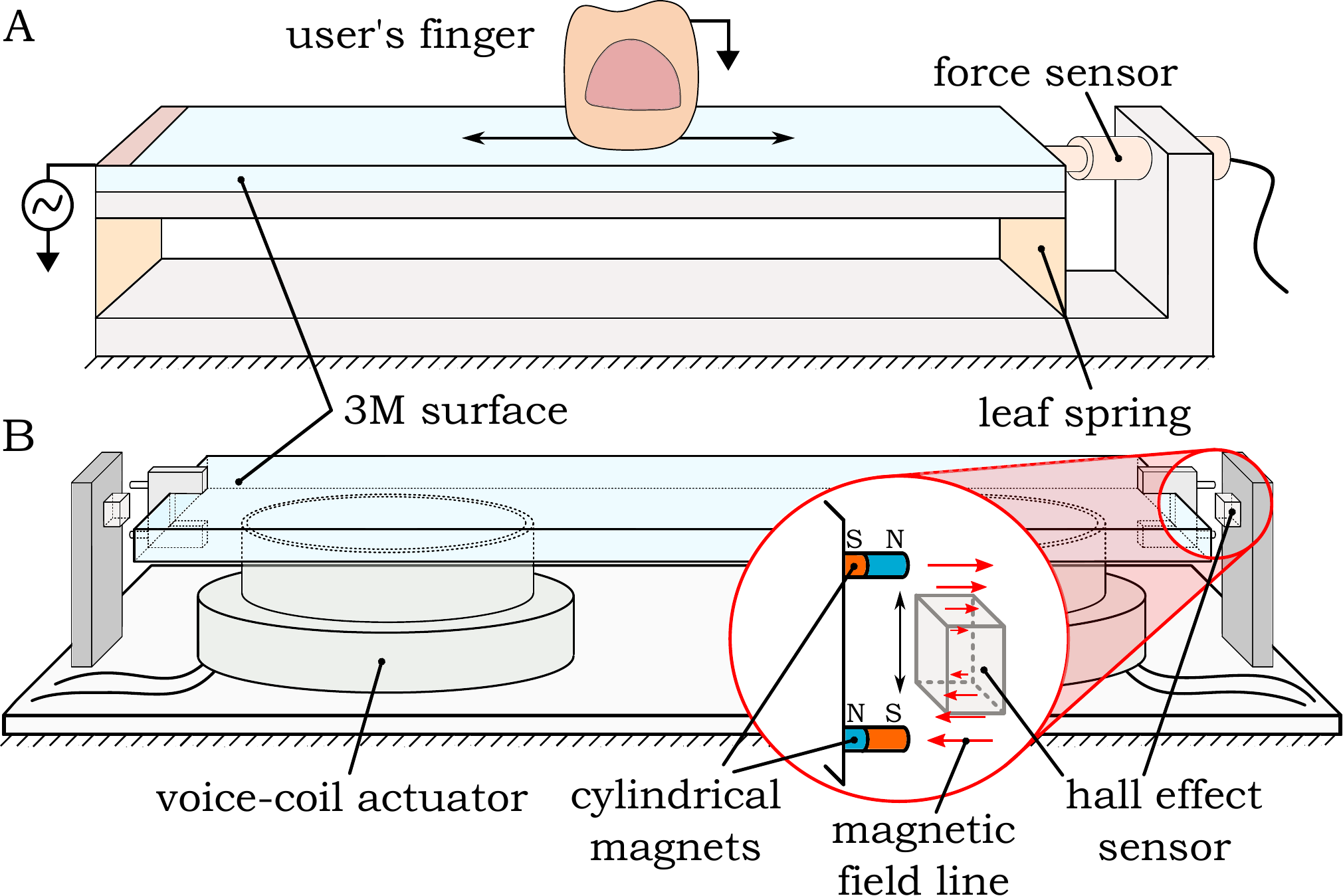}
\caption{A) Tribometer apparatus  used to generate lateral shear forces via electroadhesion based FM. B) Vibrotactile apparatus which vibrated a 3M glass using a closed-loop controller, driving two voice coil actuators in parallel.}
\label{device}
\end{figure}

A custom tribometer was built to measure high-bandwidth friction force generated between a finger and the surface of the 3M glass 25mm x 110mm in size (Figure $\ref{device}$A). A stiff piezoelectric force sensor was coupled to an aluminum frame which supported the glass and produced a flat ($\pm$ 0.5dB) impulse response in the 20-400Hz range and a clear resonance at 1.3kHz with a linearity of $R^2 \ge 99.5\%$. On this surface we generated a tactile grating by variable friction though application of electroadhesion force. The electroadhesion effect was driven by a 20kHz current carrier and modulated by: m(t) = 6$\sqrt{(\text{sin}(2\pi ft)+1)/2}$ mA where $f$ is the grating frequency. The square root operation was performed to compensate for the square law dependence of friction on current \cite{Shultz2018}. The modulated carrier was produced by a function generator (RIGOL DG1022) and converted to current by a high-bandwidth current amplifier, like the one used in \cite{Shultz2018,grigorii2020closed}.

\begin{figure}[b]
\centering
\includegraphics[width= 6 cm,keepaspectratio]{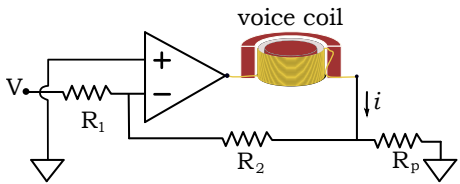}
\caption{Schematic of the current amplifier driving vibrotactile device. The current applied to a given transducer is given by $i =  \frac{V}{R_{1}}(1-\frac{R_{2}}{R_{p}})$ where $R_{p}$ is a shunt power resistor. In our case, $R_{p} = 1\Omega$, $R_{1} = 10k\Omega$, $R_{2} = 2k\Omega$ which grants $i \approx -\frac{V}{5}$.}
\label{schematic}
\end{figure}

Another 3M glass substrate of the equal size was fixed on top of two voice coil transducers (TEAX32C20-8) (Figure $\ref{device}$B). We did not aim to produce electroadhesion forces on this 3M glass. Rather, this substrate was selected to eliminate any perceptual differences between the two surfaces (e.g. temperature, overall friction, fine surface roughness) except for the ones generated due to the mode of actuation. The two voice coil actuators were separately driven by two custom transconductance amplifiers constructed from high-voltage, high-current op-amps (OPA548), as shown in Figure \ref{schematic}. The displacement of the plate relative to transducer ground was monitored by two hall effect sensors which picked up the magnetic flux generated by cylindrical magnets in a two-magnet configuration placed at each end of the glass substrate. The plate displacement to the hall-effect voltage signal exhibited a linearity of $R^{2} \ge 99.98 \%$ in the -0.5 to 0.5 mm range. The two sensors allowed for accurate sensing of surface displacement in the normal direction at the ends of the glass, and these readings were sufficient to compute normal surface displacement anywhere along the length of the stiff 3M glass substrate. 

Modulation of the electroadhesion carrier and the current in the vibrotactile apparatus was commanded by a 12bit ADC, and digitization of analog hall-effect readings was performed by a 14bit DAC. All embedded computation and communication was performed at 5kHz using a 80MHz PIC32 micro-controller. All data was sampled at 30kHz by a DAQ 60211 for offline analysis. 

\subsection{Closed loop control}
\subsubsection{Control strategy}

As discussed above, closed loop control was needed to produce the desired plate displacement in the VT apparatus. Applying such control in a traditional manner (i.e. directly over the sensed signal) would prove limiting due to the resonance of the system at a relatively low frequency ($\approx190$Hz) as well as noise associated with hall-effect sensing ($\sigma$ $\approx{1\mu}$m) which would be perceptible if reinserted into the system through control. In order to control the system beyond its open-loop bandwidth, we designed a novel controller based on application of the Short Time Discrete Fourier Transform (STDFT). The STDFT provided real-time amplitude and phase estimates which were treated as feedback signals. 

This approach relied on several characteristics of our control problem: 1) the surface needs to be driven at a discrete set of frequencies; 2) there is a linear relationship between plate displacement and its sensor reading; 3) the amplitude of the driving AC current and amplitude of transducer displacement are linearly related. The last point was verified by finding that a driving signal composed of a single frequency produced an output displacement containing harmonic distortion (ratio of harmonics to fundamental) of $<0.2\%$. We further confirmed the linearity of the current and force generated at the transducers by observing the current needed to drive a series of calibrated masses at selected amplitudes. Sufficiently linear actuation and sensing processes guaranteed that, when controlling amplitude and phase of the driving current signal at a selected frequency, most of the sensing and excitation will occur at that frequency alone.

\subsubsection{STDFT of sensed signal}

To compute amplitude and phase of plate displacement at a given frequency, we start by computing the STDFT of the digitally sampled hall-effect sensor readings, $x[n]$, at discrete time intervals, $n = \{...t_{n-2},t_{n-1},t_{n}\}$. We first multiply the signal by a wave of known amplitude (1), frequency ($f$), and phase (0): 

\begin{equation}
y_{f}[n] = x[n] \cdot e^{-j2 \pi f n/N}
\label{eq1}
\end{equation}
where $N$ is the sampling rate. The STDFT of the signal $x$ at frequency $f$, sampled and averaged across $M$ samples, can then be expressed as:

\begin{equation}
X_{f}[n] = S_{f}(x) = \frac{1}{M} \sum_{m=0}^{M-1} y_{f}[n-m]
\label{eq2}
\end{equation}
where $X_{f}$ is a complex signal that captures relative amplitude and phase information of the sensor signal over time window $M/N$ and $S_f$ refers to the STDFT operation. The time-series running average of relative amplitude and phase are respectively given by: 

\begin{equation}
    A_{f}[n] = 2|X_{f}[n]| = 2\sqrt{\mathcal{I}(X_{f}[n])^{2} + \mathcal{R}(X_{f}[n])^{2}}
    \label{eq3}
\end{equation}    
\begin{equation}
    \Theta_{f}[n] = \angle{X_{f}[n]} = \text{atan2}(\mathcal{I}(X_{f}[n]),\mathcal{R}(X_{f}[n]))
    \label{eq5}
\end{equation}
The time-averaging operation described by eq. \ref{eq2} is equivalent to applying a box-car FIR low-pass filter. This type of filter has good stability properties but a slow response, and it requires multiple digital computations to perform a low-pass operation at a small cut-off frequency relative to sampling rate (i.e. requires large M in eq. \ref{eq2}). One can replace this with an IIR low-pass filter to serve the function of computing the time-running average but with a faster overall response and smaller number of computations. Additionally, IIR filters can be represented as continuous time transfer functions which lets them be used in optimal controller design. This modified STDFT operation, henceforth referred to as STDFT* %and termed $S_{f}^{*}$
, can be expressed as:

\begin{equation}
X_{f}^{*}[n] = \sum_{m=0}^{M}  b[m]\cdot y_{f}[n-m] - \sum_{m=1}^{M}  a[m]\cdot X^{*}_{f}[n-m] 
\label{eq4}
\end{equation}
where $b$ and $a$ are the filter parameters, $M$ is the filter order (M=2 in our case), and initial values of $X^{*}_{f}$ were set to 0. The computation of amplitude and phase from the averaged signal, $X_{f}^{*}$, remains the same as it was for $X_{f}$ in eqs. $\ref{eq2}$ and $\ref{eq3}$. 

\begin{figure}[t]
\centering
\includegraphics[width=8.5 cm,keepaspectratio]{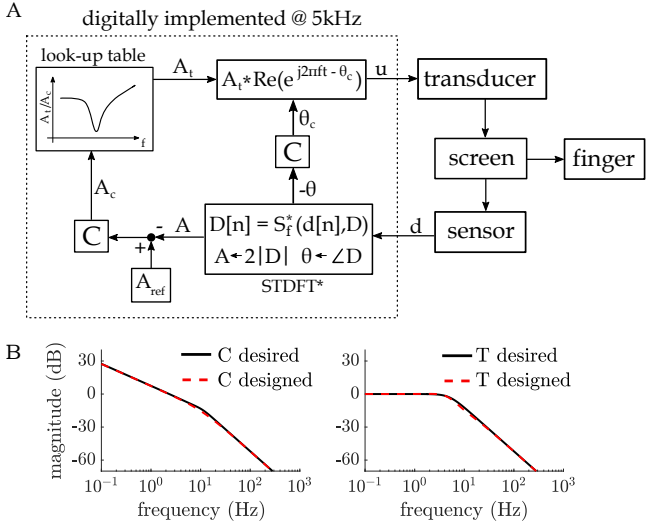}
\caption{A) block diagram of closed loop controller demonstrating the computation of signal amplitude, A, and phase, $\Theta$, at a given frequency, f, and their respective control signals, $\text{A}_{c}$ and  $\Theta_{c}$, which formed the control signal for a given transducer, u. B) The dynamics of the designed controller, C, and the resulting sensitivity function, T, of closed loop response to desired amplitude and phase.}
\label{controller}
\end{figure}

\subsubsection{Computing filter and control dynamics}

The low-pass filter order and bandwidth are limited by the constraints on the desired system bandwidth and the frequency of lowest order harmonic that we wish to close the loop over, $f_{0}$, in $x$. Through simulation, we found that, given a low-pass filter bandwidth of $\le$ $\frac{1}{2}$ $f_{0}$ and closed loop bandwidth of $\le\frac{1}{4}f_{0}$, the contribution of sensing error to the control signal will be $<$0.5$\%$ at the lowest harmonic, and will be increasingly attenuated at higher actuation frequencies. The lowest harmonic applied in our experiments was selected to be 20Hz which constrained our filter frequency cut-off to be at most 10Hz. It has been reported that finger kinematics for texture exploration largely evolve at a rate $\approx{3}$Hz \cite{callier2015kinematics}, consistent with a closed-loop bandwidth of 5Hz.

\begin{figure*}[t]
\centering
\includegraphics[width = 18 cm,keepaspectratio]{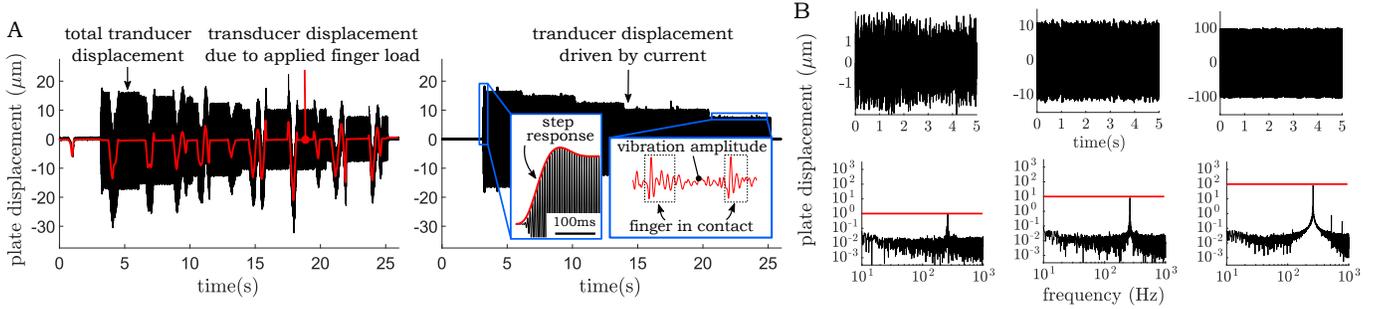}
\caption{A) Left: Total displacement generated at one end of the plate as the finger explores a vibrating surface at 261Hz. The upward displacement is generated when the finger is at the opposite end of the screen and creates a moment around a transducer. Right: Total transducer displacement band-passed to reflect plate displacement caused only by the voice coil to demonstrate control dynamics.  B) Out-of-plane surface displacement at the plate's center as it is driven at 261Hz using closed-loop control. Data was collected as the finger was actively swiping the surface while applying $\approx$ 0.5N of normal load. Top row is a temporal representation and bottom row is the spectral representation of plate vibration when the reference amplitude was set to 1,10,and 100 $\mu m$. Red line represents the desired amplitude set by GUI slider.}
\label{commanded}
\end{figure*}

The estimated amplitude and phase of the feedback signal's lowest harmonic were compared to the commanded amplitude and phase value (a user adjustable $\text{A}_{\text{ref}}$ and 0 respectively) to produce two error signals to which an optimal linear controller was applied (Figure \ref{controller}A). To that end, we employed a loop shaping technique previously adapted for surface haptic rendering \cite{grigorii2020closed}. The controller design begins by modeling the dynamic responses of each system component. In our case significant dynamics could be attributed to the transducer itself and the digital low-pass filter applied in the STDFT* computation. The former was integrated into the system as a look-up table in the frequency domain, and the latter was represented by its analytic transfer function. The closed loop response, $T$, which is termed the sensitivity function, can be represented in Laplace domain as:

\begin{equation}
T = C/(1+LC)
\end{equation}

where $L$ and $C$ are the low-pass filter and controller dynamics respectively. Selecting $T$ to be a 2$^{nd}$ order 5Hz low-pass filter and solving for $C$, we obtain:

\begin{equation}
C=T/(1-LT)
\end{equation}

This analytic solution to the optimal controller dynamics yields a 6th order transfer function. For numerical stability and computational efficiency we approximated this function with one of 2nd order (Figure \ref{controller} B). Having obtained a 2nd order solution, the discrete time parameters of the controller were extracted with Tustin's bi-linear transform. 

We found that, due to phase differences between the two voice coil actuators, the vibrating plate would exhibit rocking motion around its center of mass rather than travel up and down uniformly when only the amplitude controller was applied. This resulted in significantly more overall displacement at the two ends of the screen than in the center, especially at actuation frequencies $>$100Hz. Therefore, the phase of the driving current needed to be adjusted in real time in order to force two sides of the screen to move in synchrony. This prompted the use of closed loop controller over the phase for which we applied the same controller solution as we did for amplitude control. Since a unit change in control signal produced unit change in phase no look-up table was necessary here. Simulations showed that varying phase of driving signal will have some effect on the signal amplitude estimates, especially at or around system resonance, unless the phase control bandwidth is reduced. Through simulation and testing we confirmed that this was not a significant issue in our case, and is in general of limited concern in actuators that are sufficiently damped relative to load impedance.

The look up table in frequency reflected only the unloaded response of the system which changed when the finger came into contact with the vibrating surface. This change was especially apparent when the surface was actuated at frequencies close to resonance. Since the finger always adds impedance, the look up table will underestimate the necessary control effort and thus reduce the closed loop system bandwidth. In other words, the effect of the finger was a reduction of closed-loop bandwidth but not a loss of excitation amplitude. Through simulation we observed a 50$\%$ decrease in bandwidth when the load impedance matched that of our system. Empirically, we found the controller to behave remarkably well in mitigating error in vibration amplitude even when the apparatus was driven near resonance (Figure \ref{commanded}). 

The resulting combined control strategy showed good performance under typical applied loads and swiping speeds. Moreover, by observing the amplitude at only one specific frequency, the controller ignores irrelevant disturbances such as displacement due to the applied normal load. This grants maximum control authority over the signal of interest as can be observed in Figure \ref{commanded}A. Figure \ref{commanded}B demonstrates the controller forcing a set vibration amplitudes even when the displacement is equal to the magnitude of sensor noise, highlighting the robustness of this type of control to noisy sensing. The accuracy of actuation was thus constrained by the noise floor in the frequency domain, a sub perceptual threshold value of $0.01 \mu$m, and not the time-series value of 1$\mu$m.

\subsubsection{Additional practical considerations}

The STDFT* control contains two known sources of instability in practical application. First, since the control signal commands the amplitude of a sinusoidal signal, it must never go negative, else the feedback will become positive and grow exponentially large. Therefore a lower-bound control limit must be enforced. Second, phase computation must be carefully designed to avoid discontinuities that are the by-products of computing its relative rather than absolute value. A standard arctan function contains a discontinuity between $2^{\text{nd}}$ and $3^{\text{rd}}$ quadrants at $-180^{\text{o}}$/$180^{\text{o}}$. The phase between applied current, which is proportional to transducer force, and the resulting displacement will vary between $0^{\text{o}}$ and $-180^{\text{o}}$ as function of actuation frequency as expected given a second-order mass, spring, damper system. Thus, there is potential for discontinuity in the computed phase when the system is driven at frequencies where phase control approaches $0^{\text{o}}$ and $-180^{\text{o}}$ which will cause large, perceptible swings in control and possible instability. To avoid this we used an arctan function that experiences discontinuity at $-270^{\text{o}}$/$90^{\text{o}}$ instead by modifying the original through conditional statements, moving the computed phase discontinuity far from current/displacement phase delays exhibited by the system. 

\section{Modeling forces at periphery}
In our experiments, we ask subjects to match VT and FM stimuli in intensity.  The physical effects of these two forms of actuation are, however, somewhat different.  In the case of VT, the reaction force comes from the impedance of the musculoskeletal system.  In other words, whatever normal force the surface applies to the finger must also pass through the tissues containing the mechanoreceptors.  That normal force also modulates the friction force, which also passes through the tissues.  Thus, perception will depend on some combination of the normal and lateral forces acting on the finger.  In the case of FM, as it is achieved by electroadhesion, the reaction force comes from deformation of asperities on the stratum corneum.  As such, the applied normal force does not pass through the tissues, but the friction forces still do.  Thus, it is quite possible that the qualitative experiences of VT and FM would be distinct from one another since one includes normal force and one does not.  In practice, however, this doesn't seem to be the case, at least at the frequencies studied here. As such, we will assume that at a given frequency and equal perceived intensity the reaction force of VT, $F_r$, and additional friction force generated by FM, $F_a$, relate to one another via a simple scale factor, $\beta$.  

\section{Finger load, kinematics, and impedance}

We use a model-based approach to extract the finger's applied normal load, its location along the vibrating plate, the normal forces acting on it, and finally, its mechanical impedance. The finger's applied normal load and location were assumed to evolve slowly ($\le$10Hz) and were estimated based on low-passed displacement data and the voice coil stiffness $k$.  The expression for approximate normal load is:

\begin{equation}
W = -(k_{1}x_{1}^{l} + k_{2}x_{2}^{l})
\label{nload}
\end{equation}
where $x_{1}^{l}$ and $x_{2}^{l}$ represent low-passed normal displacements of each transducer estimated through spatial interpolation of displacement recorded at each end of the plate. Similarly, the expression for approximate finger location is: 

\begin{equation}
P = d\cdot(\frac{k_{2}x_{2}^{l}}{k_{1}x_{1}^{l}} + 1)^{-1}
\label{kinem}
\end{equation}

where $d$ is the distance between the two transducers (55 mm). Finger position and velocity data were used to constrain our impedance measurements to the middle 30mm of the substrate during steady state motion of the finger. 

We high-pass filtered plate displacement and transducer current data at 10Hz to capture all transducer borne dynamics and used them to to extract the mechanical impedance, $Z_f$, and the dynamic reaction forces, $F_f$, of the contacting finger. To estimate finger reaction forces from measurements made at the voice coil transducers, it was necessary to subtract out the reaction forces due to the VT system itself.  This was accomplished by using a set of ordinary differential equations that capture plate dynamics:

\begin{equation}
F_1 = k_{1}x_{1}^{h} + b_1\dot{x}_{1}^{h} + \frac{m}2\ddot{x}_{1}^{h} - \frac{I}{2}\ddot{\theta}^{h}
\label{eqF1}
\end{equation}

\begin{equation}
F_2 = k_2x_{2}^{h} + b_2\dot{x}_{2}^{h} + \frac{m}2\ddot{x}_{2}^{h} + \frac{I}{2}\ddot{\theta}^{h}
\label{eqF2}
\end{equation}

where $x_{1}^{h}$ and $x_{2}^{h}$ and their derivatives refer to motion at each transducer as it evolved at $\ge$ 10Hz. $\theta$ represents plate tilt angle where $\ddot{\theta}^{h} \approx (\ddot{x}_{2}^{h} - \ddot{x}_{1}^{h})/d$, the approximation being valid for small $\theta$ as is the case here owing to phase control that mitigated spurious plate rotation.

We begin solving this equation by expressing the force generated by a voice coil as function of current applied across it, namely $F = Ki$, where $i$ is current and $K$ is the transducer constant. We use the measured $m$ = 32g, $I$ = 4$\cdot$10$^{-2}$gm$^{2}$, and $K$ = 7.4 N/A parameters and time-series transducer kinematics data to find optimal parameters $k_{1}$, $k_{2}$, $b_{1}$, $b_{2}$ that fit to eqs. \ref{eqF1} and \ref{eqF2}. In practice, mechanical parameters associated with stiffness and damping of a voice coil tend to be nonlinear as well as time and frequency dependent \cite{kong2016dynamical}. The initial set of parameter values was therefore used in a regression that found optimal, first-order, time varying parameter values to fit eqs. \ref{eqF1} and \ref{eqF2} when the device was unloaded by the finger (see appendix for more detail). Optimal system parameters and measured plate motion were then used to solve for the expected $F_f$ by substituting eqs. \ref{eqF1} and \ref{eqF2} into:

\begin{equation}
F_{f} = K(i_{1} +i_{2}) - F_{1} - F_{2}
\end{equation}

Having computed $F_{f}$, the impedance associated with the finger was computed with the expression:
 
\begin{equation}
\textbf{Z}_{f}  = \frac{\textbf{F}_{f}}{\textbf{v}}
\end{equation}

\begin{figure}[t]
\centering
\includegraphics[width=8.5 cm,keepaspectratio]{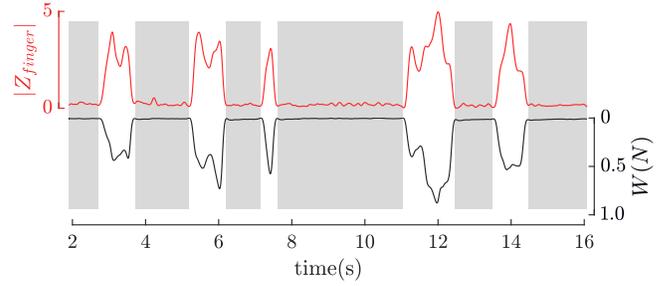}
\caption{The magnitude of finger impedance (red) as it varies in time is plotted alongside the measured applied normal load of the finger (black) during one of the experimental trials under 111Hz VT stimulation. Note that finger contact can be easily detected from normal load estimation alone which allows for segmentation of data into 'loaded' and 'unloaded' (shaded grey) regimes and for unloaded system impedance to be completely accounted for through model-driven analysis. Also note the similarity between variation in impedance and the applied normal load suggesting that dynamic properties of the fingerpad are highly load dependent.}
\label{impnor}
\end{figure}

Where the time-series functions of force and velocity were converted to phasor form to make the necessary division, i.e. $\textbf{v}$ = v$\cdot e^{-j\phi_{v}}e^{-j2\pi f t}$ where the $e^{-j2\pi f t}$ term algebraically cancels since we constrain quantities of force and velocity to the actuation frequency. For the computations discussed here, plate velocity and acceleration were obtained by performing symmetric numerical differentiation of plate displacement.  These quantities, together with force measurement, were filtered by applying a band-pass filter at frequency of actuation. The magnitude of the complex impedance was then computed as a function of time via lock-in amplification and low-passed at 10Hz, as demonstrated in Figure \ref{impnor}. 

\section{Psychophysical experiments}

Two psychophysical studies were conducted with 13 subjects (5 male, 8 female, ages = 21-32). The studies were approved by the Northwestern Univeristy IRB, subjects gave consent, and they were financially compensated for their time. The average combined experiment time was 30 minutes. Prior to the first experiment, subjects washed their hands with soap and warm water and dried them with a paper towel. They were instructed to wear active-noise cancelling headphones that played a pink noise to mask any audio cues produced by the devices (the VT device produced audible noise during use). All but one subject reported to be right-handed and all subjects used their dominant hand during judgment tasks. The subjects were not constrained in time or kinematics during the experiments.

In the first experiment, subjects were presented with a graphical user interface (GUI) containing a slider that they used to adjust the amplitude of sinusoidal plate displacement. In an earlier pilot study it was found that the velocity of the plate vibration was the best overall predictor of perceived vibration intensity. The GUI slider was therefore reset at every trial to a range between 0 and the displacement amplitude which corresponded to maximum velocity of .06m/s (e.g., 0.48mm displacement amplitude at 20Hz and 0.024mm at 400Hz). This gave the subjects the ability to utilize most of the slider range in an adjustment task. Subjects were instructed to freely explore the two surfaces by medial-lateral motion while adjusting the VT slider until the perceived intensity of vibration matched that of the FM device. The matching task was performed at eight logarithmically spaced frequencies: $\{$20.0, 31.0, 47.0, 72.0, 111.0, 170.0, 261.0, 400.0$\}$ Hz, four times per frequency, totaling 32 trials per subject. Presentation of the stimuli were pseudo-randomized to present each frequency once during trial numbers 1-8, 9-16, 17-24, and 25-32. 

In the first block of eight trials (corresponding to the 8 distinct frequencies), the slider was set to 0. During the second block, the slider was initialized to values 50\% greater than those chosen during the first. In the third and fourth blocks, the slider value was initialized to -50\% and +50\% of the thus-far expected selection of VT displacement at the given frequencies. This approach ensured that the subjects converged to a perceptual match from both directions, giving us the option to compute the unbiased mean of the matched VT stimulus as well as the standard deviation around it. 

In the second experiment we asked the subjects to make similarity judgments between VT and FM stimuli when both were driven at the same frequency with VT plate displacement set at the average intensity match found in the first experiment. Subjects responded using a slider that ranged from "not at all similar" to "one and the same". This experiment took place immediately after the first and was composed of eight trials - one trial per frequency.

\begin{figure}[t]
\centering
 \includegraphics[width = 8.5 cm,keepaspectratio]{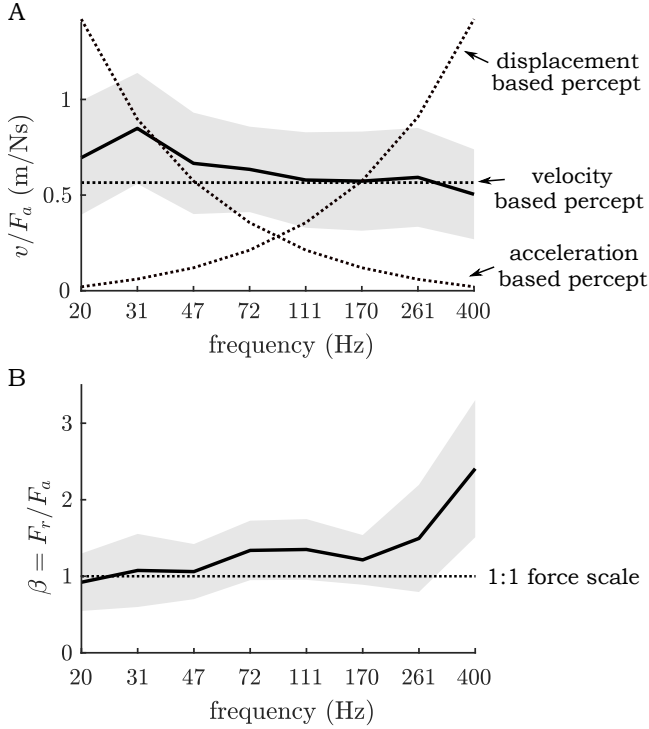}
\caption{A) Mean ratio (black) of the user-selected plate velocity, $v$, to the experienced FM force, $F_{a}$, when the two stimuli were perceptually matched, with standard deviation represented as grey shade. B) Ratio between the VT normal reaction force, $F_{r}$, and the FM lateral force, $F_{a}$.}
\label{combinedvel}
\end{figure}

\section{Results}
\subsection{Perceptual domain in intensity matching}

Figure \ref{combinedvel}A plots the ratio of VT plate velocity to FM friction force at the point of equal perceived intensity.  This  ratio is mostly flat across stimulation frequencies, with an average value of $\approx 0.64$ m/Ns. The inverse of this ratio, which constitutes the effective perceptual intensity based `damping', is consistent with the value of mechanical damping of skin in shear (1.6Ns/m reported here vs. 1.5Ns/m found in \cite{wiertlewskimechanical}). The positive slope at the far left end of the plot suggests that, at low frequency, VT plate displacement rather than velocity may best correlate to FM lateral force.  Here, the ratio constitutes an effective perceptual intensity based `stiffness' of $\approx$ 150N/m, consistent with the reported value of finger stiffness that approximately falls in the range of 100-250N/m \cite{hajian1997,pawluk1999dynamic}. 

Figure \ref{combinedvel}B plots $\beta$, the ratio of VT normal reaction force to FM lateral force at the point of equal perceived intensity.  Clearly, $\beta$ depends on actuation frequency (t-test = 6.4, p$<10^{-8}$), growing at frequencies above 170Hz.  At 47Hz and below, $\beta$ is close to unity, suggesting that, at low enough frequency, perception may depend on the overall magnitude of applied force, irrespective of direction or mechanism.

\begin{figure}[t]
\centering
\includegraphics[width= 6 cm,keepaspectratio]{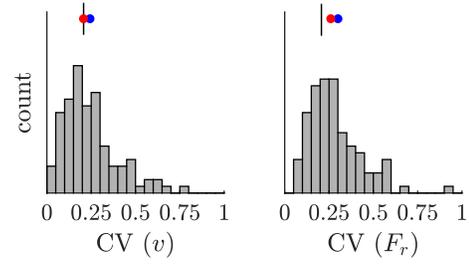}
\caption{Coefficient of variance (CV) for selected velocity and experienced reaction force. Red and blue points represent medians and means respectively, black vertical lines are aligned to one another for reference.}
\label{COV}
\end{figure}

Nonetheless, the overall greater flatness of $v/F_{a}$ compared to that of $\beta$ over the full frequency range implicates the motion of the plate as the dominant perceptual intensity parameter in VT. We wished to further validate the motion model against the force model by observing how sensitive subjects were to changes in plate motion versus force in VT stimulation. A measure of perceptual sensitivity to a given physical stimulus parameter in a method of adjustment experiment is the coefficient of variation (CV), computed as CV = $\sigma/\mu$, where $\sigma$ and $\mu$ represent the standard deviation and mean of the selected stimulus value for a given subject at a given frequency. We computed CV for subject-adjusted plate velocity and reaction force and pooled all computed CV values together (Figure \ref{COV}). The CV value for $F_{r}$ (median = 0.275) was 30\% greater than that for $v$ (median = 0.205); moreover, the difference in two distributions was statistically significant (2 sample t-test (n$_1$ = 90, n$_2$ = 92) = 2.57, p<10$^{-1}$). Thus, it appears that subjects were generally more sensitive to changes in $v$ when making adjustments to VT stimulus.

\subsubsection{Variability of peripheral excitation}

\begin{figure}[b]
\centering
\includegraphics[width= 8.5 cm,keepaspectratio]{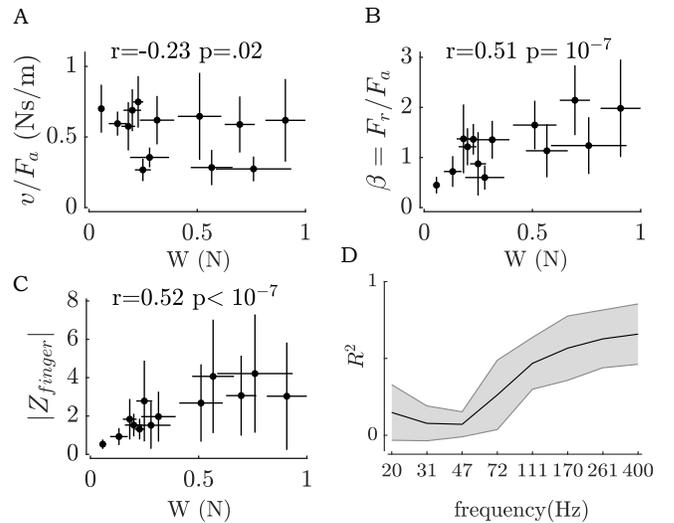}
\caption{A) Selected VT velocity to FM shear force ratio. B) VT reaction to FM shear force ratio $\beta$. C) Finger impedance plotted against the normal applied load for every subject. Black points and lines represent means and standard deviations across trials for a given subject. D) coefficient of determination between W and $Z_{f}$ during a given swipe across actuation frequencies.}
\label{ZBv}
\end{figure}

We found a strong dependence of the fingertip impedance and the reaction force on the applied normal load, $W$, as is apparent in Figure \ref{impnor}. The magnitude of impedance correlated with $W$ during a given swipe (r = .66, p < $10^{-10}$), implying large variability of generated force to touch conditions during a any trial, and across all subjects (r = 0.85, p = $2\cdot10^{-4}$). In other words, the forces generated at the periphery across the subject population were largely determined by how hard they pushed on the surface. Notably, the VT velocity, $v$, which was selected to match $F_{a}$ in intensity was not significantly affected by the applied normal load (r=-.13, p = 0.17) as can be observed in Figure \ref{ZBv}A, unlike the impedance and the reaction force (Figure \ref{ZBv}B-C). We found that this observed variability in impedance can be largely attributed to variability in normal load at frequencies $\ge$111Hz and not lower frequencies, Figure \ref{ZBv}D. No similar dependence of $v$ on $W$ exists due the efforts of the controller. It appears that dynamic qualities of the finger in that frequency range are most sensitive to normal load. Power analysis revealed that $|Z_f| \propto W^{1/3}$ at this frequency range as well, in line with Hertzian contact theory \cite{wiertlewskimechanical} of skin. It appears that the variability of finger impedance is tied to mechanics of the skin and tissue. We thus have some mechanical evidence in support of the earlier observation of perceptual preference for $v$ over $F_{r}$ especially in the upper frequency range. 

\subsection{Modeling fingertip impedance in touch}

\begin{figure}[t]
\centering
\includegraphics[width= 8.5 cm,keepaspectratio]{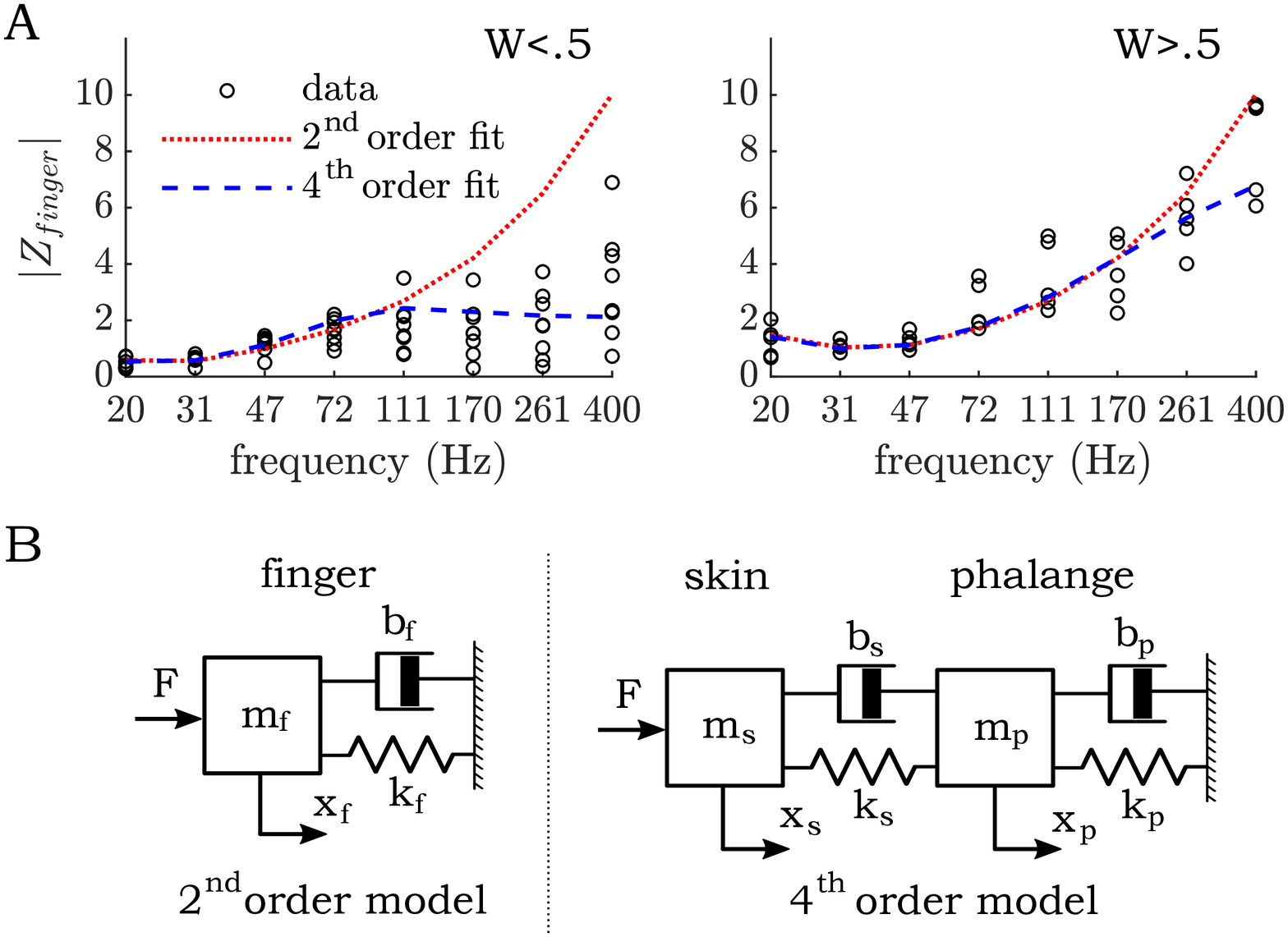}
\caption{A. Computed fingertip impedances extracted from data collected during the psychophysical experiment B. Two dynamic models used to fit the above data, the left lumping skin/tissue and phalanx parameters together, and the right treating them separately.} 

\label{combinedimpedance}
\end{figure}

We developed a dynamic model of the interaction between the finger and the vibrating plate and fit it to our data. First, however, we segregated the impedance measurements into two groups, one where subjects applied $<$0.5N of normal load and one where they applied $>$0.5N of load during their perceptual task (Figure \ref{combinedimpedance}A). It can be immediately observed that the impedance data collected at W$>$0.5N are well-predicted by a second order model of the fingertip (parameters m$_{f}$, b$_{f}$, k$_{f}$ shown in Table 1 were taken from \cite{hajian1997}). In contrast, data collected for subjects that applied relatively smaller loads, W<0.5, could not be fit well with a second order model at the higher frequencies, regardless of parameter values.

\begin{table}[t]
\begin{center}
\renewcommand{\arraystretch}{1.5}
  \begin{tabular}{ | c | c | c || c | c | c | }
      \hline
    \multicolumn{3}{| c ||}{W<0.5} 
    &\multicolumn{3}{ c |}{W>0.5}\\ \hline
    \hline
    m$_{f}$  & b$_{f}$ & k$_{f}$ & m$_{f}$  & b$_{f}$  & k$_{f}$ \\ 
    \hline
    4.0 & 1.0 & 0.2 & 4.0 & 1.5 & 0.3 \\ 
    \hline 
    \hline
    m$_{p}$  & b$_{p}$ & k$_{p}$ & m$_{p}$  & b$_{p}$  & k$_{p}$  \\ 
    \hline
    4.0 & 1.0 & 0.2 & 4.0 & 1.5 & 0.3 \\ 
    \hline
    \hline
    m$_{s}$  & b$_{s}$ & k$_{s}$ & m$_{s}$  & b$_{s}$  & k$_{s}$ \\ 
    \hline
    0.2 & 1.5 & 1.0 & 0.2 & 8.0 & 8.0 \\ 
    \hline
  \end{tabular}   
\end{center}
\caption{Parameter values used for impedance fits. Units for m,b,and k are in g, N$\cdot$s/m, and N/mm respectively.}
\label{table1}
\end{table}

Impedance data collected under light touch were instead fit with a fourth order model that lumped mechanical parameters of the skin (and underlying tissue) and those of the phalanx separately as depicted in Figure \ref{combinedimpedance}B. For this fit, finger phalanx impedance parameters ($b_p,k_p$) were kept the same as those found for the whole finger and the skin impedance parameters ($m_s,b_s,k_s$) were initially adapted from \cite{wiertlewskimechanical}. Their values are presented in Table \ref{table1}. The 4$^{th}$ order model fit both sets of impedance data better but appeared to capture dynamic behavior under low applied loads especially well. We found, however, that in order to fit impedance data at larger applied loads, the stiffness and damping parameters of skin needed to be increased by a factor ranging from 2 to 8 from those reported in shear, the exact factor depending on $W$. This finding suggests that skin stiffness and damping in the normal direction are load-dependent and may be altogether larger than values observed in shear. This finding compares well with published finger impedance measurements \cite{wiertlewskimechanical,pawluk1999dynamic}. 

One implication for the current study is that force generation at the periphery is sensitive to the direction of stimulation and much more so at higher frequencies where the skin tends to stiffen rapidly with applied load. Additionally, the success of a second order VT model at low frequency suggests that forces and motions of the skin will match those of the phalanx as a whole.  In other words, the motion of the bone will track that of the vibrating surface.  In contrast to this, the necessity to model skin as a separate dynamic entity to fit high frequency impedance data suggests that the bone fails to track the surface at those higher frequencies. This is most obvious at lower applied loads since at low enough skin stiffness the mass of the finger fails to be dynamically engaged.  

\subsection{A proposed perceptual mechanism}

\begin{figure}[t]
\centering
\includegraphics[width= 8.5 cm,keepaspectratio]{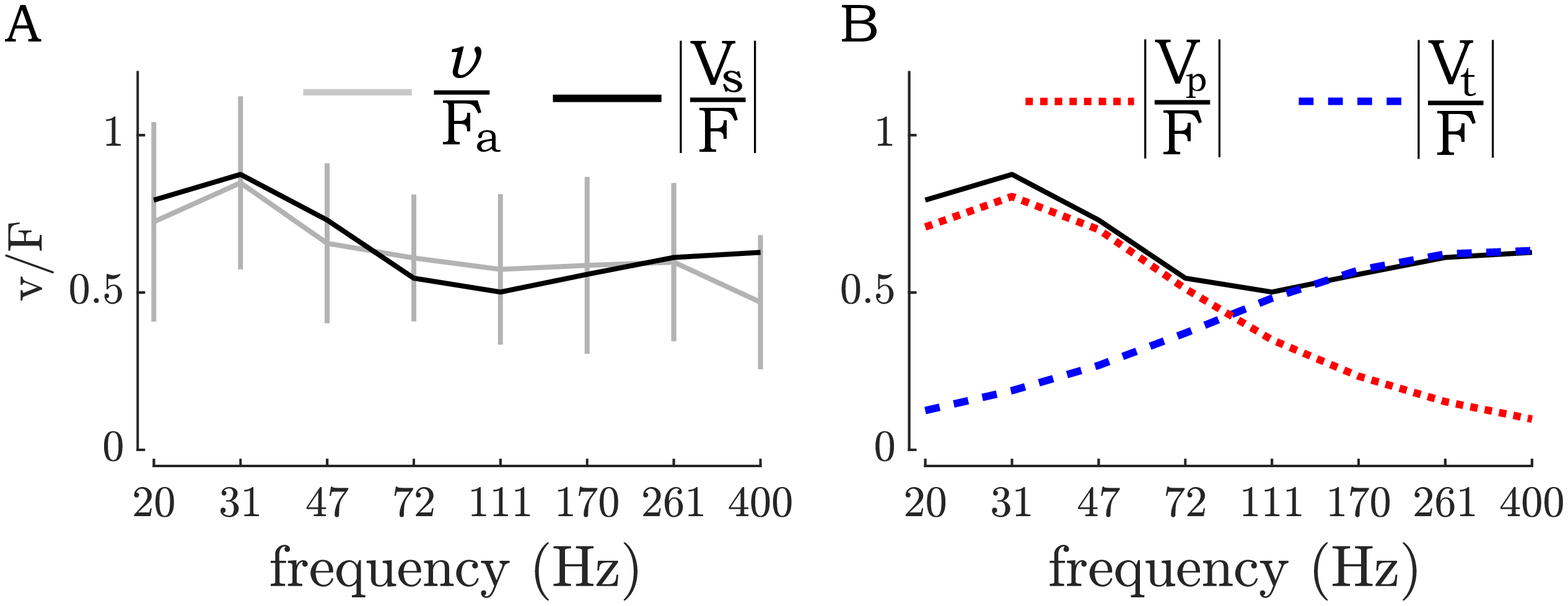}
\caption{A) Magnitude of lateral skin velocity to the applied shear force ratio as one would expect in FM stimulation, modeled by the 4th order model (black). Normal skin velocity achieved in VT set-up divided by the shear force achieved in FM (grey). B) Magnitude of velocity to force ratio, where force is applied at periphery and the velocity is computed at the tissue (dashed blue) and phalanx (dotted red), plotted alongside the admittance curve from A).}
\label{vF}
\end{figure}

In this section, we appeal to the dynamic model of Figure \ref{combinedimpedance} to help explain the perceptual results of Figure \ref{combinedvel}.  To begin, the black line in Figure \ref{vF}A is the shear admittance magnitude computed from the 4$^{th}$ order model using the same parameters borrowed from literature and used to fit our impedance measurements.  On the same axes, shown in grey, is the ratio of the intensity-matched normal plate velocity $v$ in VT stimulation to the shear force in FM stimulation.  The marked similarity of the two curves suggests that equivalent intensities occur when the magnitude of the skin velocity, regardless of direction, is matched.  Note that no fitting is done here, we simply apply the data available in literature with reasonable modifications to account for specifics of touch conditions. 

Although fingertip velocity may explain perceived intensity, it is unlikely that it can be associated with a single set of sensory receptors.  This is illustrated by plotting the transfer functions from applied force to both phalanx velocity and skin velocity (relative to the phalanx) as shown in Figure ref{vF}B (refer to eqs. \ref{phalanxmob},\ref{tissuemob} in the appendix).  The sum of these two transfer functions is the admittance shown in Figure \ref{vF}A.  It is apparent that, at low frequency, most of the fingertip motion arises from the phalanx, while at high frequency, most of the motion arises from the skin.  It would stand to reason that two different receptor populations -- one at the joint \cite{edin1991finger} and one in the skin \cite{johansson1976skin} -- would be involved as well.

\subsection{Similarity judgments}

\begin{figure}[b]
\centering
\includegraphics[width = 8.5 cm,keepaspectratio]{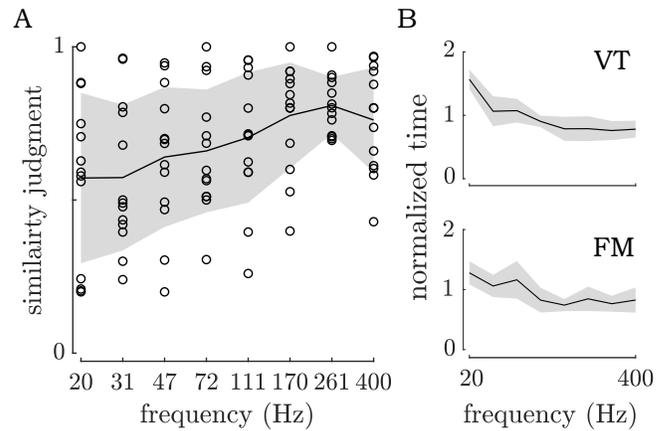}
\caption{A) Similarity rating across all subjects between VT and FM stimulation. B) Contact time needed in the intensity matching task normalized by the average time each subject spent in their match task. }
\label{subres}
\end{figure}

After performing the VT vs. FM intensity matching task the average velocity each subject selected at a given frequency was used to create a VT grating which was then compared in similarity to that of FM based grating generated at the same electroadhesion current as in previous experiment. The results of this experiment are shown in Figure \ref{subres}A. The reported similarities were found to be proportional to actuation frequency in the 20-400Hz range (t-test(104) = 4.03, $p<10^{-4}$) with a $42\%$ increase in similarity from 20Hz to 261Hz. The peak in similarity occurs at 261Hz, confirming our earlier hypothesis that the two types of gratings would feel most similar at higher frequencies where direction of stimulation is not encoded at the mechanoreceptor level. Interestingly, 261Hz also marks the frequency of least spread in similarity judgment. The similarity judgments slightly decrease at 400Hz relative to 261Hz with a larger overall spread in choices as well, but it is not clear why. We observed no significant relationship between the similarity rating and $v$, $F_{r}$, or $F_{a}$, suggesting that the similarity rating did not correlate with stimulus intensity. Thus, the goal of removing intensity-based cues in the similarity judgment appears to have been successful. 

\subsection{Emergent exploration tactics}

Some aspects of exploratory procedures varied across actuation frequencies in a manner consistent with our proposed mechanism of perception. Figure \ref{subres}B demonstrates the time spent exploring each surface during the intensity matching task, normalized by average time taken by each subject at a given trial. Low frequency stimuli were explored nearly twice as long for both FM (t-test(104) = 3.5, p$<10^{-3}$) and VT (t-test(104) = 6, p$<10^{-7}$). This hints at a different type of perceptual mechanism at low frequency.  For instance, an exploration time inversely proportional to frequency may be evidence of integration (accumulation) of information, such as counting the number of grating periods experienced across a swipe. We found a weak but significant negative correlation between exploration time and average scan velocity in support of the latter notion ($\rho$ = -.27 p$<$.01).

We found no significant difference between exploration times of FM and VT surfaces, and no changes in surface exploration strategies, i.e. scanning speed, pressing force, were observed based on stimulation type or grating frequency.

\section{Discussion}

The results reported here strongly suggest that, across a broad frequency range, gratings induced by normal displacement of the touch surface and those induced by lateral force modulation are perceived as being equivalent in intensity when the magnitude of velocity at the fingertip matches, irrespective of vibratory direction.  At sufficiently low frequency ($<$72Hz), it is also true that equivalence in intensity occurs when the forces on the fingertip match, irrespective of direction.  This appears to stem from one, the finger and skin moving in concert; and two, the finger impedance being dominated by damping \cite{wiertlewskimechanical}, which maps force directly to velocity.  At higher frequencies, as the dynamics of the fingertip tissue come into play, the force match breaks down, but the velocity match does not.  It is remarkable that intensity remains so tightly coupled to fingertip velocity despite the fact that motion shifts from predominantly joint-level at the lower frequencies to predominantly skin-level at the higher frequencies.

We observed large variations in fingertip impedance under variable applied normal loads, $W$, which led to variations of peripheral reaction force, $F_{r}$, making it an objectively poor predictor of relative VT intensity at high actuation frequencies. Moreover, it appeared that subjects were less sensitive to $F_{r}$ than $v$, preferring the latter when making intensity choices. 

Finally, we found a considerable amount of similarity between the two actuation types which tended to improve with increasing frequency.  This result is consistent with perception being a blend of direction-sensitive kinesthetic sensing and tactile sensing at the lower frequencies, transitioning to primarily spatially and direction-insensitive (e.g. PC afferent \cite{johansson1976skin,abraira2013sensory}) tactile sensing at the higher frequencies.  This result has implications for the design of future haptic interfaces.  For instance, the addition of low-bandwidth normal-direction VT to high-bandwidth FM may enhance realism by providing greater control over directionality at the low frequencies.

\subsection{Extended application of STDFT based control}

The STDFT control presented here may be applied to any haptic application that relies on actuation at discrete frequencies to produce a desired tactile effect. One of major benefits of this approach lies in its relative insensitivity to system dynamics since only the dynamics at frequencies close to carrier are relevant to control, thus offering potential for signal control across a narrow band or a across a series of bands that compose a wide-band signal. The latter point gives STDFT control a potential utility in wide-band playback of texture features, such as those generated by VT, FM, or other forms of stimulation with limited sensitivity to noncolocation, nonlinearities, or sensing noise.   

STDFT control may be utilized in indirect modes of tactile stimulus that do not rely on force sensing. For example, ultrasonic plate vibration amplitude, which is physically more predictive of friction reduction than driving piezo voltage\cite{vezzoli2017friction}, can be controlled for more accurate friction modulation. Since ultrasonic devices are driven by a carrier of at least 20kHz the control bandwidth can be set a large fraction of it (according to the math detailed earlier). In addition, a measurement of phase between the driving and motion signals offers the potential for locking the ultrasonic carrier frequency to the system resonance, where it is most energetically efficient and offers the largest dynamic range. In electrostatic systems, STDFT control could be utilized in sensing electrical impedance of the interface gap that has been strongly implicated in generated friction force \cite{Shultz2018}, thus modulating the current to produce a desired electrostatic force. These applications can lead to overall  higher quality rendering of rich tactile stimuli. 
\section{Acknowledgements}
This work was supported by the National Science Foundation grant number IIS-1518602.  The authors would like to thank Dr. Roberta Klatzky for many productive discussions.

\section{Appendix}
\subsubsection{Modeling system parameters}

The damping and stiffness parameters of each voice-coil transducer were found to be time-dependent. A reasonably good fit to these parameters could be achieved via a first order time-variable model, e.g. $k,b$ = $a\cdot t$ + $c$. The time varying component of system parameters was small relative to the base values, i.e. $c \gg a$, nevertheless, it proved necessary to account for them to accurately compute the impedance of the finger. Initial values of time based model-parameter were set to $a$ = 0 and $c$ was set to the initially reported or measured values of $k_{1}$ = 22.4 kN/m, $k_{2}$ = 22.9 kN/m, $b_{1}$ = $b_{2}$ = 0.85 Nm/s, during regression.

\subsubsection{Fingertip impedance}

Using the fingertip models presented in Figure \ref{combinedimpedance}B, we derived the force balance equations. Treating the finger as a $2^{nd}$ order system with mass, damping, and stiffness parameters $m_f, b_f,$ and $k_f$ the governing force balance equation becomes:

\begin{equation}
F = m_f \ddot{x}_s + b_f\dot{x}_s + k_f x_s
\end{equation}

where $x_{s}$ and its time derivatives refer to displacement of skin at the periphery and $F$ is the reaction force acting on the finger. The associated impedance at periphery of the skin is:

\begin{equation}
Z_{s} = \frac{m_f s^2 + b_f s + k_f}{s}
\label{2ndorderZ}
\end{equation}

When treating the phalanx and skin/tissue of the finger as separate dynamic entities where $m_p,b_p,$ and $k_p$ are phalanx parameters and $m_s,b_s,$ and $k_s$ are skin parameters we obtain $4^{th}$ order force balance equations:

\begin{equation}
F_{r} = m_s \ddot{x}_s + k_s(x_s - x_p)+ b_s(\dot{x}_s - \dot{x}_p) \\
\end{equation}
\begin{equation}
0 = m_p\ddot{x}_p + b_p\dot{x}_s + k_p x_p + b_s(\dot{x}_p - \dot{x}_s) + k_s(x_p - x_s)
\end{equation}

where $x_{s}$ and $x_{p}$ and their time derivatives refer to motion of skin at periphery and phalanx, respectively. Solving for impedance at the periphery of skin:

\begin{equation}
Z_s = \frac{A}{m_p s^3 + (b_p + b_s)s^2 + (k_p + k_s)s} 
\label{4thorderZ}
\end{equation}

\begin{gather*}
\text{where}: \;\;\; A = (m_s m_p)s^4 + (m_s(b_p + b_s) + m_p b_s)s^3\;+\\ 
(m_s (k_s + k_p) + m_p k_s + b_s b_p)s^2+(b_s k_p + b_p k_s)s  + k_s k_p \\
\end{gather*}

\subsubsection{Deriving velocity of skin and phalanx}

We now compute the velocity of the skin and that of the phalanx due to applied peripheral force. In the case of skin, this quantity is simply the inverse of skin impedance:

\begin{equation}
\frac{V_s}{F} = \frac{1}{Z_s} = \frac{m_p s^3 + (b_p + b_s)s^2 + (k_p + k_s)s}{A}.
\label{skinmob}
\end{equation}

Phalanx velocity due to input peripheral force is:

\begin{equation}
\frac{V_p}{F} = \frac{b_s s^2 + k_s s}{A}
\label{phalanxmob}
\end{equation}

The velocity of skin tissue relative to the phalanx (i.e., the bone), can thus be found from these two relations:

\begin{equation}
\frac{V_{t}}{F} = \frac{s(X_s -X_p)}{F} = \frac{m_p s^3 + b_p s^2 + k_p s}{A}  
\label{tissuemob}
\end{equation}

\bibliography{bib.bib} 

\vspace{5cm}

\begin{IEEEbiography}
    [{\includegraphics[width=1 in,keepaspectratio]{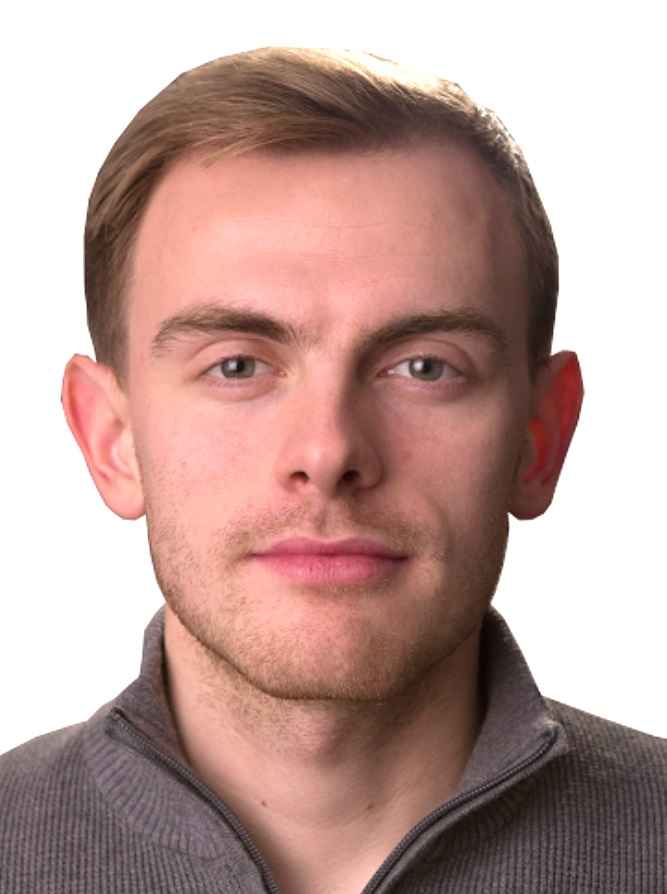}}]{Roman V. Grigorii} received his BA degree in Physics from Grinnell College and is currently a PhD candidate in Mechanical Engineering at Northwestern University. His research interests lie in development of tools for capture, subsequent playback, and enhancement of tactile cues as they are rendered directly to the human finger. One of his primary research goals is to understand the physical aspects of touch that aid perception of surface features such as textures and how they contribute to realism.  He is a member of IEEE. 
\end{IEEEbiography}    

\begin{IEEEbiography}
    [{\includegraphics[width=1in,clip,keepaspectratio]{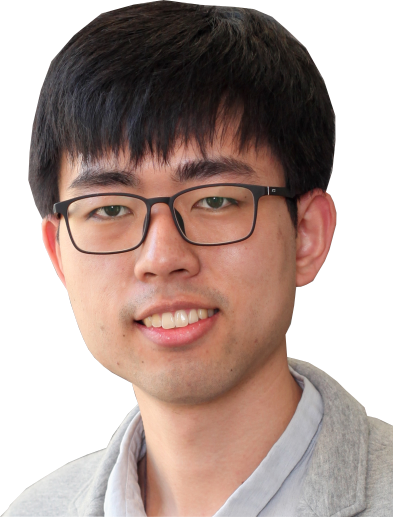}}]{Y. Evan Li} received his Master's degree from the Department of Mechanical Engineering at Northwestern University in 2019. He is currently working at a startup based in Shenghai China, where he focuses on integrating haptics technology into virtual reality applications.
\end{IEEEbiography}

\begin{IEEEbiography}[{\includegraphics[width=1in,height=1.25in,clip,keepaspectratio]{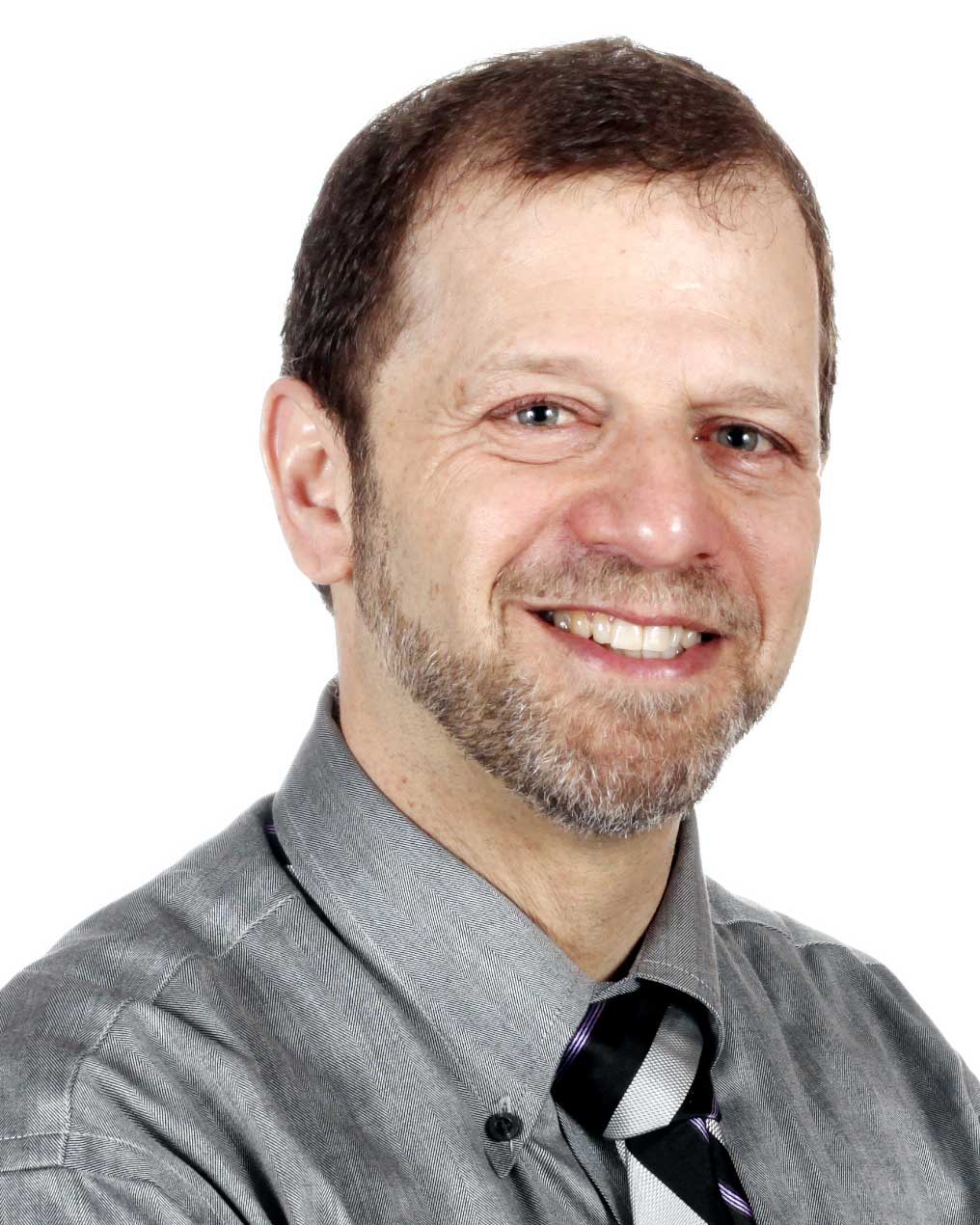}}]{Michael A. Peshkin}
 (SM’09) is a professor in the Department of Mechanical Engineering at Northwestern University (Evanston, IL.)  His research is in haptics, robotics, human–machine interaction, and rehabilitation robotics. He has cofounded four start-up companies: Mako Surgical, Cobotics, HDT Robotics, and Tanvas. He is a Fellow of the National Academy of Inventors, and (with J. E. Colgate) an inventor of cobots.
\end{IEEEbiography}

\begin{IEEEbiography}
    [{\includegraphics[width=1in,clip,keepaspectratio]{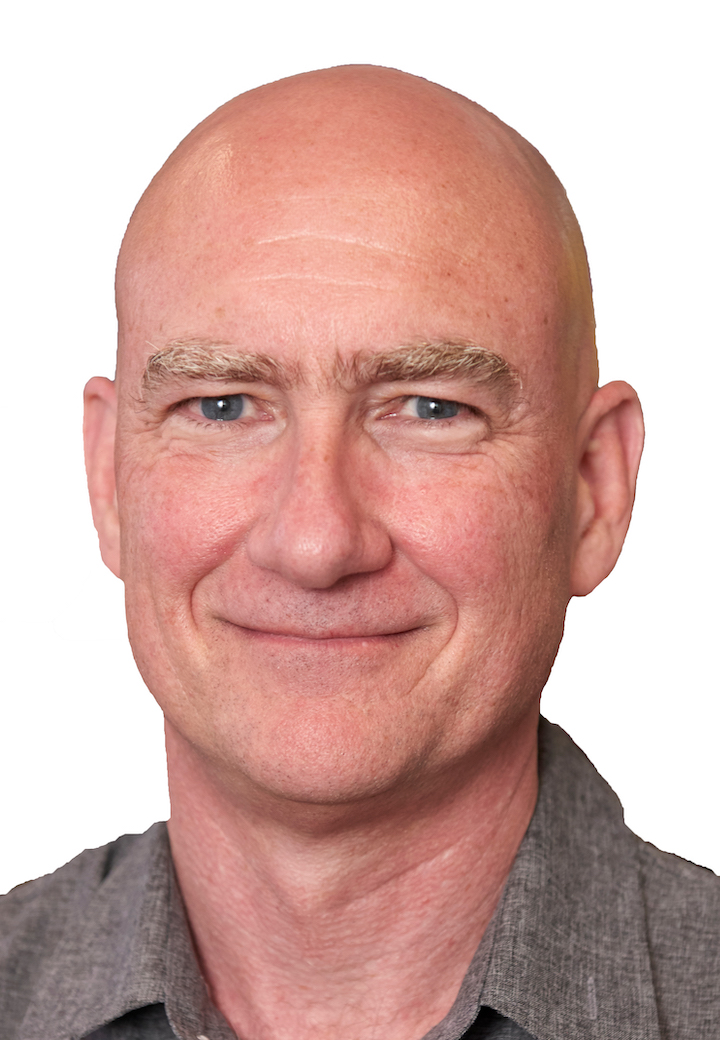}}]{J. Edward 
    Colgate} is a professor of Mechanical Engineering at Northwestern University in Evanston, Illinois.  Dr. Colgate's principal research interests are haptic interfaces and human-robot 
    interaction. Dr. Colgate was the founding Editor-in-Chief of the IEEE Transactions on Haptics, a Fellow of the IEEE and the National Academy of Inventors, and a member of the Chicago Area Entrepreneurship Hall of Fame. Dr. Colgate was one of the founding co-directors of the Segal Design Institute at Northwestern University where he directed the Master of Science in Engineering Design and Innovation, which combines graduate-level engineering courses with a broad exposure to human-centered design.  Colgate and collaborator Michael Peshkin are the inventors of cobots, which led to their first startup together, Cobotics Inc. Their second startup, Kinea Design, developed advanced physical therapy robots and prosthetic limbs.  Their third company together, Tanvas Inc., is commercializing innovative surface haptic technologies that allow users to feel tactile effects on a touch screen.
\end{IEEEbiography}

\end{document}